\documentclass[showpacs,aps,prb,twocolumn,superscriptaddress,10pt]{revtex4-2}
\usepackage{graphicx} 
\usepackage{bm}
\usepackage{color}
\usepackage{amsmath}
\usepackage{amssymb}
\usepackage{enumerate}
\usepackage{xspace}
\usepackage{mathrsfs}
\usepackage{mathptmx}
\usepackage[normalem]{ulem}
\usepackage[colorlinks=true,linkcolor=blue,citecolor=blue,urlcolor=blue]{hyperref}
\newcommand{\be}{\begin{equation}}
\newcommand{\ee}{\end{equation}}
\begin{document}
\title{The Influence of Finite-size Effects on the Curie Temperature of L1\textsubscript{0}-FePt}
\author{Nguyen Thanh Binh}
\affiliation{%
Department of Physics, the University of York, Heslington, York YO10 5DD, United Kingdom}%
\author{Sergiu Ruta}
\affiliation{%
Department of Physics, the University of York, Heslington, York YO10 5DD, United Kingdom}%
\author{Ondrej Hovorka}
\affiliation{%
Faculty of Engineering and Physical Sciences, the University of Southampton, Southampton
SO17 1BJ, United Kingdom}%
\author{Richard. F. L. Evans}
\affiliation{%
Department of Physics, the University of York, Heslington, York YO10 5DD, United Kingdom}%
\author{Roy. W. Chantrell}
\affiliation{%
Department of Physics, the University of York, Heslington, York YO10 5DD, United Kingdom}%
\begin{abstract}
We employ an atomistic model using a nearest-neighbor Heisenberg Hamiltonian exchange to study computationally the dependence of the Curie temperature of $L1_0$-FePt on finite-size and surface effects in Heat-assisted Magnetic Recording (HAMR) media. We demonstrate the existence of a size threshold at 3.5nm below which the impact of finite-size effects start to permeate into the center of the  grains and contribute to the reduction  of the Curie temperature. We find a correlation between the Curie temperature and the percentage of atomistic bonds lost on the surface as a function of grain size, which can be extended to apply to not only $L1_0$-FePt but also generic magnetic systems of any crystal structure. The investigation gives insight into finite-size effects which, because the inevitable grain size dispersion leads to an irreducible contribution to a dispersion of the Curie temperature, has been predicted to be a serious limitation of HAMR.
\end{abstract}
\maketitle
\thispagestyle{empty}

\section{INTRODUCTION}
\label{S:1}
Magnetic recording functions on the balance between three main factors which form the well-known magnetic trilemma: the task of optimizing the signal-to-noise (SNR) ratio, thermal stability, and writability~\cite{Richter2007}. To improve areal density while maintaining sufficient SNR, the grain volume in the recording medium is reduced leading potentially to a loss of thermal stability. Hence, the requirement is to find a recording layer material with high uniaxial anisotropy energy density $K$ to ensure that the thermal stability factor $KV/kT\ge 60$ with $V$ being the grain volume. However, using a high-anisotropy material for the recording medium produces yet another issue: that a high magnetic field from the writing transducer would be required to switch the grain magnetization. In principle, a fourth factor has to be taken into account: a probability of back-switching of spins during the assisted-writing process due to thermally induced transitions. Currently this acts as a source of DC noise, however  in terms of ultra-high storage densities involving heated dot recording this gives a potential limit of magnetic recording density~\cite{Richard1}. 

\begin{figure}[!ht]
\centering
 \includegraphics[angle = -0,width = 0.4
 \textwidth]{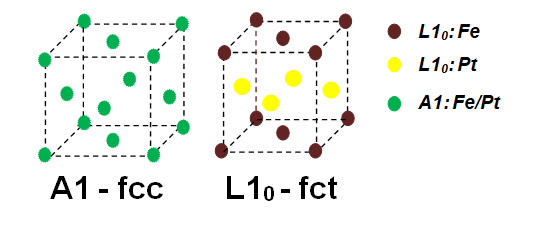}
   \caption[]{Crystal structures of FePt: (left) disordered A1-fcc bulk-alloy $Fe_{0.5}Pt_{0.5}$ at room temperature; (right) ordered $L1_0$-fct at HAMR temperature}
\label{fig_FePtphases}
\end{figure}

Recently, Heat-assisted Magnetic Recording (HAMR) has emerged as a solution to circumvent the problem presented by the magnetic trilemma~\cite{Rottmayer2006,Mcdaniel2005}. The HAMR writing head first applies an intensive, highly localized heat spot for a very short time to a recording medium to heat it up to or beyond its Curie temperature ($T_C$), then writes the data inductively after which cooling to ambient temperature restores the thermal stability. Among many possible candidates for a HAMR recording layer material, Iron Platinum in $L1_0$-phase ($L1_0$-FePt) has been regarded as an excellent choice~\cite{Kryder}. $L1_0$-FePt has been widely studied for application in HAMR in which $L1_0$-FePt can function either as a single layer or as part of a composite multi-layer recording medium~\cite{Weller1,Loc,Thiele,Barucca,Zhou,Wang}. FePt undergoes a transition at around 300\textsuperscript{o}C from the bulk alloy A1 phase  with a disordered, randomly distributed face-center cubic (fcc) crystal structure to a chemically ordered face-center tetragonal (fct) crystal structure in the $L1_0$ phase~\cite{Nakaya}. In the $L1_0$ phase, FePt comprises of alternating layers of 3d-element Fe and 5d-element Pt atoms along the (001) direction - as sketched in Fig.~\ref{fig_FePtphases}. In the $L1_0$ phase the Fe spins polarise the Pt spins whose large spin-orbit coupling  results in the very high magnetic uniaxial anisotropy necessary for the thermal stability of written information~\cite{Mryasov,Kryder}. 

Simulations by Li and Zhu~\cite{Li2013,Zhu2014} have shown that the dispersion of $T_C$ is a serious limitation for the ultimate storage density achievable for HAMR. Consequently, a crucial aspect for successful HAMR media is controlling the Curie temperature dispersion of the recording medium. However, the exact Curie temperature of $L1_0$-FePt has yet to be established; rather, it has been reported to fall between 650K and 780K under various treatments and measurements~\cite{Waters2019}. Also, $L1_0$-FePt was shown to exhibit a strong dependence on grain size~\cite{Lyberatos,Hovorka}. Consequently, in a recording medium using $L1_0$-FePt the grain size distribution which always exists would inevitably lead to an irreducible dispersion of Curie temperature $\sigma_{T_c}$ which potentially limits the recording density. Therefore, it poses an important question to determine the precise dependence of the Curie temperature distribution in $L1_0$-FePt grains on finite size effects as well as the governing mechanisms behind it. 
 
In this paper, we present a computational investigation of the impact of finite-size effects in $L1_0$-FePt grains using a nearest-neighbor Heisenberg Hamiltonian atomistic spin model in which short-range exchange interactions are assumed to dominate. Although the exchange interactions in FePt are long-ranged, Waters et al.~\cite{Waters2019} have shown that the critical exponent of FePt conforms to the three dimensional Heisenberg universality class. Our starting point is the consideration of thermal fluctuations, which can modify $T_C$ and even shift the ground state solutions to induce changes in magnetic phases. To see this, one needs to look at how the free energy changes as a function of system size. Generally, the free energy will have a functional part reflecting the surface effects due to reduced coordination number, but there will be another term that will correspond to renormalization of state energies resulting from fluctuations~\cite{goldenfeld2018lectures}. Which one is more important will depend on the system and the size. We hypothesize a correlation between the Curie temperature distribution of $L1_0$-FePt and the percentage of atomistic bond loss on the surface of the grains as a function of grain size, which allows to separate the surface and fluctuation contributions to the finite-size effects. We find that our hypothesized correlation could be extended to encompass the role of crystal structures, which suggests it is not restricted to $L1_0$-FePt specifically but universally applicable for any generic magnetic system.

\section{THEORIES}
\label{S:2}
Here we describe the two main theories used in our calculation: the Atomistic Spin model and the Mean-field model.

\subsection{Atomistic Spin Model}
Numerical simulations are first carried out using an Atomistic Spin model. The energy of a magnetic system is generally described in terms of the Hamiltonian $\mathcal{H}$ as the sum of all energy contributions. The three  most important contributors include: the exchange interaction between pairs of local spins $\mathcal{H}_{exchange}$, the magnetic uniaxial anisotropy $\mathcal{H}_{anisotropy}$, and the externally applied magnetic field $\mathcal{H}_{field}$. The explicit form of these terms in the Hamiltonian are given as follows:
\be
    \mathcal{H} = -\frac{1}{2}\,  \sum_{i,j } J_{ij}\, (\hat{\mathbf{s}}_i \cdot  \hat{\mathbf{s}}_j)-k_u \sum_{i} (\hat{\mathbf{s}}_i \cdot  \hat{\mathbf{e}})^2 - \sum_{i} \mu_i  (\hat{\mathbf{s}}_i \cdot  \mathbf{B}),
\ee
where $J_{ij}$ is the exchange energy strength between $\hat{\mathbf{s}}_i$ and $\hat{\mathbf{s}}_j$ - the unit vector of local spin at site $i$ and $j$ respectively, $k_{u}$ the uniaxial anisotropy constant having an easy direction along $\hat{\mathbf{e}}$, $\mu_{i}$ the atomic spin moment, and $\mathbf{B}$ the externally applied magnetic field.

The Hamiltonian of $L1_0$-FePt in principle includes energy contributions from both Fe and Pt components. However, Mryasov et al.~\cite{Mryasov}, using Density Functional Theory in the constrained local-spin-density approximation, theoretically demonstrated that the Hamiltonian of $L1_0$-FePt could be rewritten in terms of Fe degree of freedom only. Using this approach, the original fct-structured $L1_0$-FePt in Fig.~\ref{fig_FePtphases} can be represented as a simple cubic structured Fe-only system with modified lattice properties - a new configuration henceforth referred to as the modified sc-FePt. An implication immediately follows that it would then be possible to extend the scope of our study to incorporate the role of different crystal structures beyond the original fct $L1_0$-FePt.

\subsection{Lattice Site Resolved Mean-field Model}
Penny et~al.~\cite{meanfield2019} have shown that a mean-field approach is valuable for the investigation of finite-size effects, including lattice types and particle shapes. To support the interpretation of our atomistic model calculation, we use a lattice site resolved Mean-field model outlined as follows~\cite{meanfield2020}. Consider the standard Heisenberg spin Hamiltonian including an applied field $\mathbf{B}$ and here using for convenience the same spin notation as~\cite{meanfield2020}:
\be
\label{spinham}
\begin{split}
    \mathcal{H} = -\frac{1}{2}J\sum_{\langle ij\rangle}\hat{\mathbf{s}}_i\cdot\hat{\mathbf{s}}_j -\mu\sum_i\hat{\mathbf{s}}_i\cdot\mathbf{B},
\end{split}
\ee
where the individual terms represent the ferromagnetic exchange interaction energy and the Zeeman energy. The symbol $\langle\cdot\rangle$ in the first sum implies that only the nearest neighbor spin pairs are summed over. The spin variables are unit vectors $\hat{\mathbf{s}}_i = \bm{\mu}_i/\mu$, $i = 1,\dots, N$, where $\bm{\mu}_i$ is the magnetic moment associated with the spin $i$ and $\mu = |\bm{\mu}_i|$ is its magnitude.

A conventional way to derive the mean-field approximation is to express the spin variables in Eq.~\eqref{spinham} as $\mathbf{s}_i = \mathbf{\tilde m}_i + \delta\mathbf{s}_i$, where $\mathbf{\tilde m}_i$ and $\delta\mathbf{s}_i$ are respectively the thermally averaged and fluctuating parts of the spin variable $\mathbf{s}_i$. Neglect the fluctuations $\delta\mathbf{s}_i$ beyond the first order and rewrite Eq.~\eqref{spinham} as:
\be
\label{spinhammf}
\begin{split}
    \mathcal{H}_\mathrm{mf} = \frac{1}{2}J\sum_{\langle ij\rangle}\mathbf{\tilde m}_i\cdot\mathbf{\tilde m}_j -\sum_i\mathbf{\hat s}_i\cdot\left(J\sum_{j\in i}\mathbf{\tilde m}_i+\mu\mathbf{B}\right),
\end{split}
\ee
where the expression in the parentheses:
\be
\label{beff}
\begin{split}
    \mu\mathbf{B}_i^\mathrm{e} = J\sum_{j \in i}\mathbf{\tilde m}_j + \mu\mathbf{B}
\end{split}
\ee
is the effective field acting on the mean-field spin moment $\mathbf{\tilde m}_i$ due to its neighbors $j$ and is derived as the variational derivative with respect to $\mathbf{\tilde m}_i$. The notation $j\in i$ means the summation is carried out over all interacting neighbors $j$ of the spin $i$. The mean-field spin moment $\mathbf{\tilde m}_i$ can be evaluated from Eq.~\eqref{spinhammf} using the canonical statistical mechanics:
\be
\label{mimf}
    \mathbf{\tilde m}_i = \frac
    {\mathrm{Tr}_{\mathbf{s}_i}\mathbf{s}_i\exp{\left(-\beta\mathcal{H}_\mathrm{mf}\right)}}
    {\mathrm{Tr}_{\mathbf{s}_i}\exp{\left(-\beta\mathcal{H}_\mathrm{mf}\right)}}.
\ee
Note that since $\mathbf{\tilde m}$ is no longer a unit vector, Eq.~\eqref{spinhammf} no longer conforms with the usual Heisenberg definition of the exchange. However, if we transform to unit vectors by multiplying through by $|\mathbf{\tilde m}|^2/|\mathbf{\tilde m}|^2$ we are left with the prefactor of the summation term in Eq.~\eqref{spinhammf} as $J\mathbf{\tilde m}^2$ which represents the temperature dependence of the effective exchange in the mean-field sense~\cite{Atxitia2010}. Upon considering that stable moment configurations are aligned with their effective fields, i.e. $\mathbf{\tilde m}_i\parallel \mathbf{\tilde B}_i^\mathrm{e}$, allows expressing $\mathbf{\tilde m}_i$ as:
\be
\label{lanf}
    \mathbf{\tilde m}_i = \mathcal{L}\left(\beta\mu|\mathbf{\tilde B}_i^\mathrm{e}|\right)\frac{\mathbf{\tilde B}_i^\mathrm{e}}{|\mathbf{\tilde B}_i^\mathrm{e}|}.
\ee
Here $\mathcal{L}(x) = \coth x - x^{-1}$ is the Langevin function, and $\beta = (k_BT)^{-1}$ with $k_B$ being the Boltzmann constant and $T$ the temperature. Equations Eq.~\eqref{beff} and \eqref{lanf} represent a set of coupled nonlinear algebraic equations which can be solved iteratively in a straightforward way, as discussed elsewhere~\cite{meanfield2020}.

\section{RESULTS AND DISCUSSION}
\label{S:3}
We have investigated the temperature-dependent magnetic properties of a material with the Curie temperature of FePt as a function of the grain size. Our simulations are carried out using the VAMPIRE atomistic simulation software package~\cite{VAMPIRE,Richard2}. We construct parallelepiped FePt grains with a nominal constant height (z) of 10nm and square base of variable size (x,y) in a nominal range from 1nm to 10nm in 0.5nm increment. The unit cells of all configurations of FePt, following Mryasov et al.~\cite{Mryasov}, have dimensions of $x_0=y_0=2.72$\r{A} and $z_0=3.85$\r{A}. To avoid creating incomplete unit cells on the grain surface, the grains are made to comprise of integer numbers of unit cells in each dimension, so the exact xyz sizes of each grain are multipliers of $x_0$ closest to their corresponding nominal values. 

\begin{figure}[!ht]
\centering
 \includegraphics[angle = -0,width = 1.0
 \linewidth]{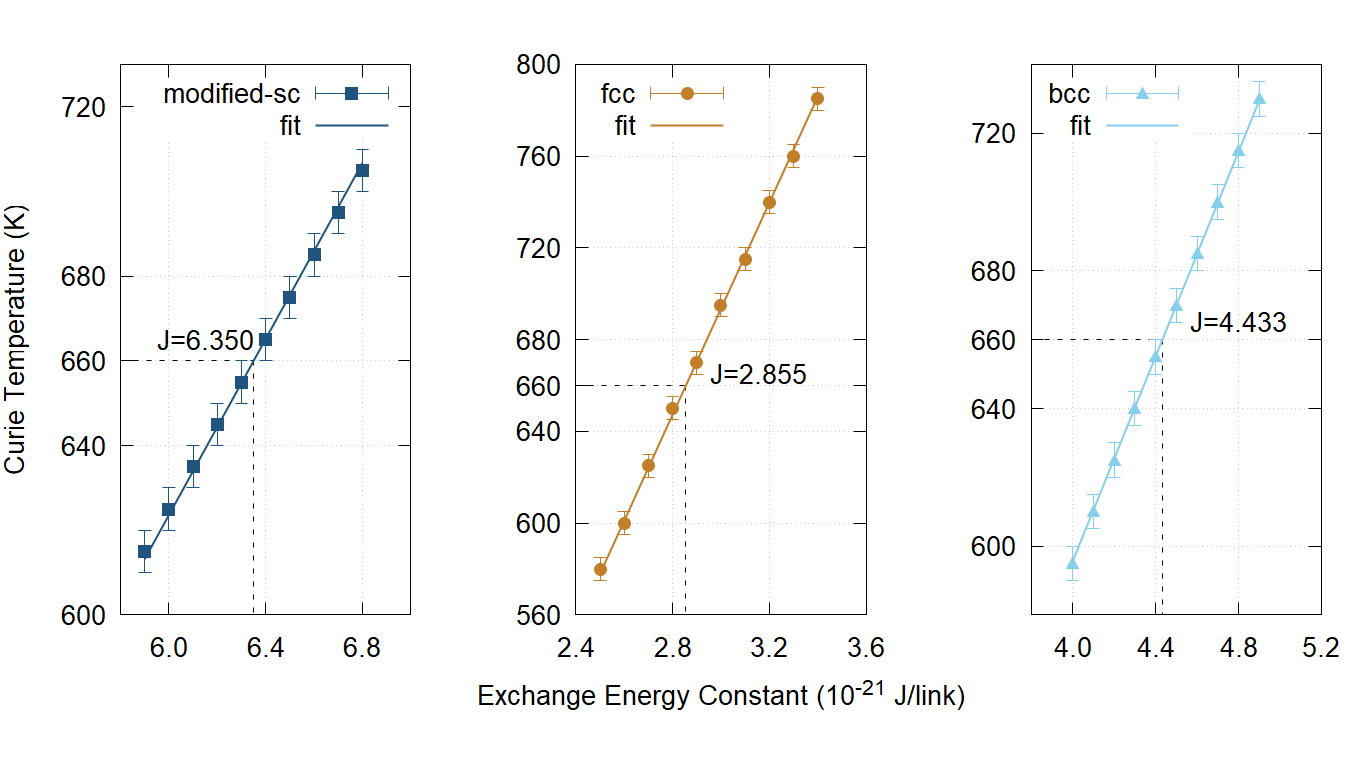}
   \caption[]{Determination of the exchange energy constant for grains with different lattice structures. As expected there is a linear dependence of Curie temperature on exchange strength. By interpolation we determine the exchange energy to give a consistent $T_C=660$K for each lattice structure. }
\label{fig_ExchangeEnergy}
\end{figure}

As aforementioned, three lattice structure configurations are simulated: a modified-sc configuration representing the original fct $L1_0$-FePt, and two "artificial" fcc and bcc FePt configurations which for comparison purpose are made to share the same magnetic attributes of the original fct $L1_0$-FePt. The effects of crystal lattices are investigated with the Curie temperature in each case first preset to a theoretically calculated critical temperature of 660K~\cite{Kazantseva} in the largest grain of 10nm base size. The Curie temperature $T_C$ is linearly dependent on the  exchange energy strength $J_{ij}$~\cite{Garanin} given by:
\be
\label{Jlinear}
    J_{ij} = \frac{3k_BT_C}{\epsilon z},
\ee
where $k_B$ is the Boltzmann constant, $z$ the number of nearest-neighbor interactions in a unit-cell, and $\epsilon$ the correction factor relating to the coordination-dependent spin wave stiffness. Note that Eq.~\eqref{Jlinear} can also be derived from the Mean-field theory, as shown in the Supplemental section S1. Both $z$ and $\epsilon$ are uniquely determined for each crystal structure~\cite{Richard2,Garanin}. Since the exchange interaction in our simulations is set to involve only nearest neighbors, the exchange interaction strength $J_{ij}$ between each pair of nearest neighbors can reasonably be assumed to be similar. Therefore, $J_{ij}$ can henceforth be simplified to the exchange energy constant $J$. For a specific lattice structure, $J$ is determined computationally by interpolation to give a consistent $T_C=660$K for each lattice structure using Eq.~\eqref{Jlinear} as shown in Fig.~\ref{fig_ExchangeEnergy}. Numerical values of $J$, $z$, and $\epsilon$ are given in Table~\ref{tab_ParametersUnique}. Other parameters representing magnetic properties of FePt which are shared by all three simulated crystal lattice configurations include the atomic spin moment $\mu_S$ and uniaxial anisotropy constant $k_u$, which are fixed at $3.23\mu_B$ and $2.63 \times 10^{-22}$ Joule/atom respectively~\cite{PhysRevApplied.14.014077}.

Our VAMPIRE simulations use a Metropolis Monte-Carlo integrator to compute the value of normalized magnetization distribution M(T) at a given  temperature T. The Curie temperature is computed from the normalized mean magnetization length and mean susceptibility $\chi(T)$ of the grain as functions of temperature. The grain size-dependent Monte-Carlo (MC) parameters used in VAMPIRE simulations are given in Table~\ref{tab_MCsteps}. Simulations are repeated 20 times to compute statistical values.  

\begin{center}
\begin{table}[ht!]
\centering
\fontsize{8}{6}
\begin{tabular}{c|llll}
Configuration & z & $\epsilon$ & J (Joule per link) \\ 
\hline \hline 
modified-sc & 6 & 0.719 & $(6.303 \pm 0.004)  \times 10^{-21}$ \\
fcc & 12 & 0.790 & $(2.866 \pm 0.002) \times 10^{-21}$ \\
bcc & 8 & 0.766 & $(4.430 \pm 0.002) \times 10^{-21}$ \\ 
\hline
\end{tabular}
\caption[]{Unit cell parameters for each simulated configuration of FePt.}
\label{tab_ParametersUnique}
\end{table}
\end{center}
\begin{center}
\begin{table}[ht!]
\centering
\fontsize{8}{6}
\begin{tabular}{c|lll}
Grain Size (nm) & Equilibration step & Total time step 
\\ \hline \hline 
1.0 - 1.5 & $5 \times 10^{7}$ & $2.5 \times 10^{8}$ \\
2.0 - 2.5 & $10^{7}$ & $10^{8}$ \\
3.0 - 5.5 & $10^{6}$ & $10^{7}$ \\
$\geq$ 6.0 & $10^{5}$ & $10^{6}$ \\ \hline
\end{tabular}
\\
\caption[]{VAMPIRE MC parameters}
\label{tab_MCsteps}
\end{table}
\end{center}

The Curie temperature variation with grain size $T_C$(D) is described by the Finite-size Scaling Law (FSSL)~\cite{Waters}:
\be
\label{Eq4}
    T_{C}(D) = T_{C}(bulk)(1-x_{0}D^{-1/\nu}),
\ee
where $D$ is the characteristic grain size, $\nu$ is a critical exponent and $x_0$ is a fitting parameter on the order of the lattice spacing. Equation Eq.~\eqref{Eq4} is applied to determine $x_0$ and $\nu$ as well as the bulk Curie temperature $T_C$(bulk). The percentage Curie temperature decrease, ${\Delta}T_{C}(D)$, can then be defined as the percentage difference between the Curie temperature at each grain size $T_{C}(D)$ and the bulk Curie temperature $T_{C}(bulk)$ obtained from the FSSL fit:
\be
\label{FSSL}
    {\Delta}T_{C}(D) = \frac{T_{C}(bulk)-T_{C}(D)}{T_{C}(bulk)}
    = x_{0}D^{-1/\nu}.
\ee
It is important to note that fits to the FSSL often give values of $\nu$ which differ: in apparent contradiction to the nature of $\nu$ as a universal exponent. As a result, $\nu^{-1}$ is often replaced by a so-called shift exponent $\lambda$ which may or may not agree with $\nu$ depending on the various system properties. Here we will propose a possible mechanism for the deviation from the critical exponent.

\begin{figure}[ht!]
\centering
 \includegraphics[angle = -0,width = 1.0 \linewidth]{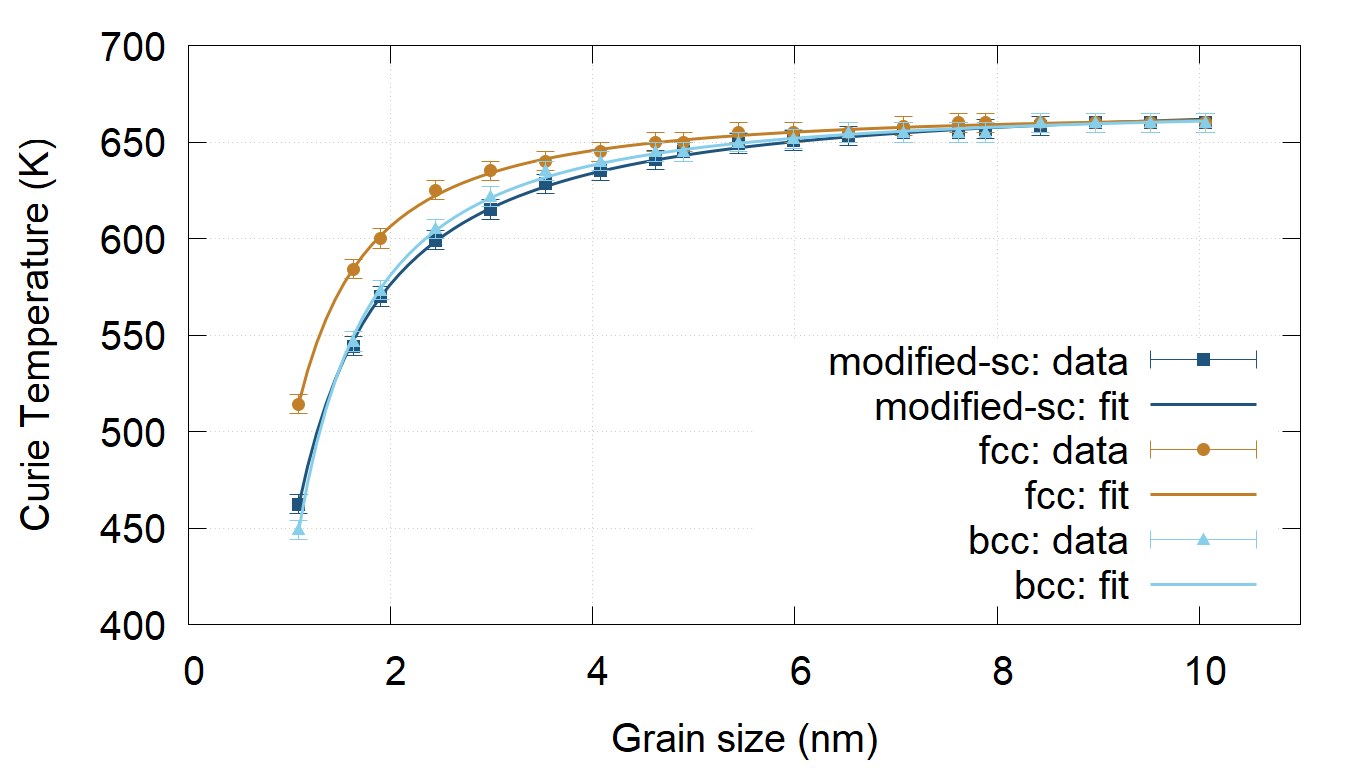}
   \caption[]{The dependence of Curie temperature on grain size fits well with the FSSL for all 3 simulated lattice structures. The $T_C$ increases sharply at smallest grain sizes and converges to the bulk value from around 4nm.}
\label{fig_TcDistribution}
\end{figure}

We proceed to an investigation of the finite-size dependence of the Curie temperature using the atomistic model outlined earlier. Fig.~\ref{fig_TcDistribution} demonstrates that the dependence of the Curie temperature on size $T_C$(D) fits well to the Finite-size Scaling Law (FSSL)~\cite{Waters} given in Eq.~\eqref{Eq4}, consistent with previous experimental data~\cite{Hovorka}. Numerical values extracted from the FSSL for the critical exponents $x_0$ and $\nu$ as well as the bulk Curie temperature $T_C$(bulk) are given in Table~\ref{tab_FSSL}. We note that, although the FSSL fits the data well for all grain sizes investigated, the values found for $\nu$ do not agree with the expected value of $0.7$ for the Heisenberg model~\cite{Hovorka,Yeomans}. Let us consider how the FSSL is obtained. It follows from the correlation scaling relation $\xi\sim (1-T/T_c)^{-\nu}$~\cite{Aharoni} which leads to the scaling law, Eq.~\eqref{FSSL}. Note that this scaling relation for $\xi$ is valid only for bulk systems, leading, in fitting to experimental data, to the empirical replacement of the exponent $\nu$, which is universal, by the shift exponent $\lambda$ which is non-universal and may include corrections to scaling. Here we consider the possibility that, for small grain sizes, the correlation length becomes dependent on the characteristic size as a result of the rescaling of system energies resulting from the fluctuations.

\begin{center}
\begin{table}[ht!]
\centering
\fontsize{8}{6}
\begin{tabular}{c|llll}
Configuration & $T_{C}(bulk)$ & $x_{0}$ & $\nu$
\\ \hline \hline 
modified-sc & $674.2 \pm 1.1$  & $0.352 \pm 0.002$ & $0.783 \pm 0.014$  \\
fcc & $666.2 \pm 0.7$ & $0.260 \pm 0.002$ & $0.651 \pm 0.013$  \\
bcc & $668.3 \pm 0.9$ & $0.375 \pm 0.003$ & $0.656 \pm 0.011$ \\ 
\hline
\end{tabular}
\caption[]{FSSL fitting parameters for each simulated configuration of FePt.} 
\label{tab_FSSL}
\end{table}
\end{center}

\begin{figure}[ht!]
\centering
 \includegraphics[angle = -0,width = 1.0 
 \linewidth]{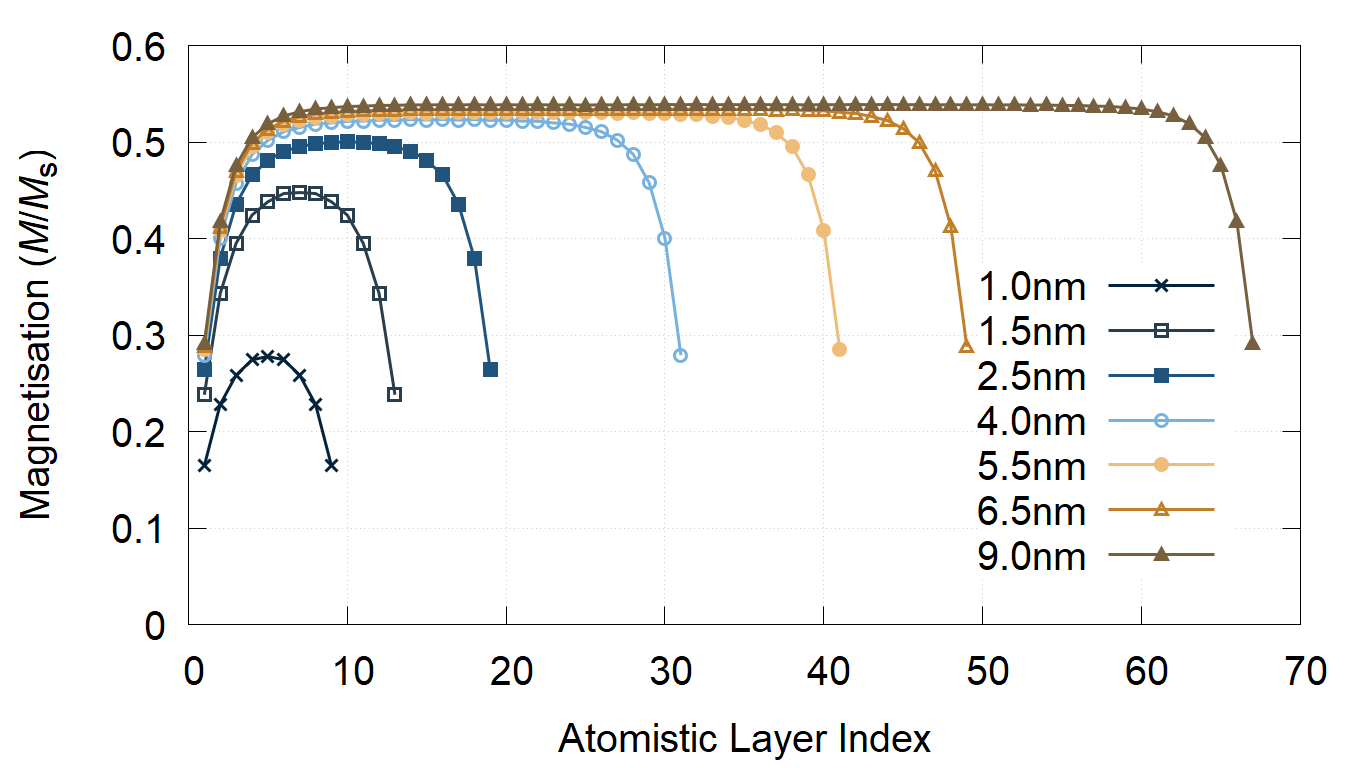}
   \caption[]{The layer-resolved magnetization profile for fcc lattice grains of selected sizes at 550K showing that in smaller grains (fewer atomistic layers) the magnetization drop on the surface contributes more to the overall loss of the grain magnetization.}
\label{fig_MagnetisationProfile-fcc}
\end{figure}

To provide further insight we have carried out a detailed analysis of the layer-resolved magnetization profiles which are obtained by averaging the spins in each layer. Examples of the magnetization profile at 550K in the x-direction along the grain depth are given in Fig.~\ref{fig_MagnetisationProfile-fcc} for different grain sizes. In larger grains (9.0nm) surface disorder (low magnetization) causing the drop in magnetization is seen to penetrate only a few layers inside the grain. On the contrary, in smaller grains where the total number of layers is reduced to the 10-12 range, surface effects begin to dominate. Our calculation shows that in smaller-sized grains the loss of order causing the decrease of Curie temperature across surface layers propagates into the center of the grain: an effect potentially responsible for causing a larger overall drop in the Curie temperature of the whole grain. Here we present results for the fcc lattice: results for modified-sc and bcc lattices are shown in Supplemental section S2. In S2-Fig.1(b) it is shown that there is a periodic behavior for the bcc lattice, which is a physical effect arising from atoms having different numbers of nearest neighbors. Interestingly this persists to elevated temperatures in the bulk of the grain albeit somewhat reduced at the grain boundaries, supporting the idea that the disorder propagates inward from the surfaces.

\begin{figure*}[htbp!]
\centering
 \includegraphics[angle = -0,width = 1.0 
 \linewidth]{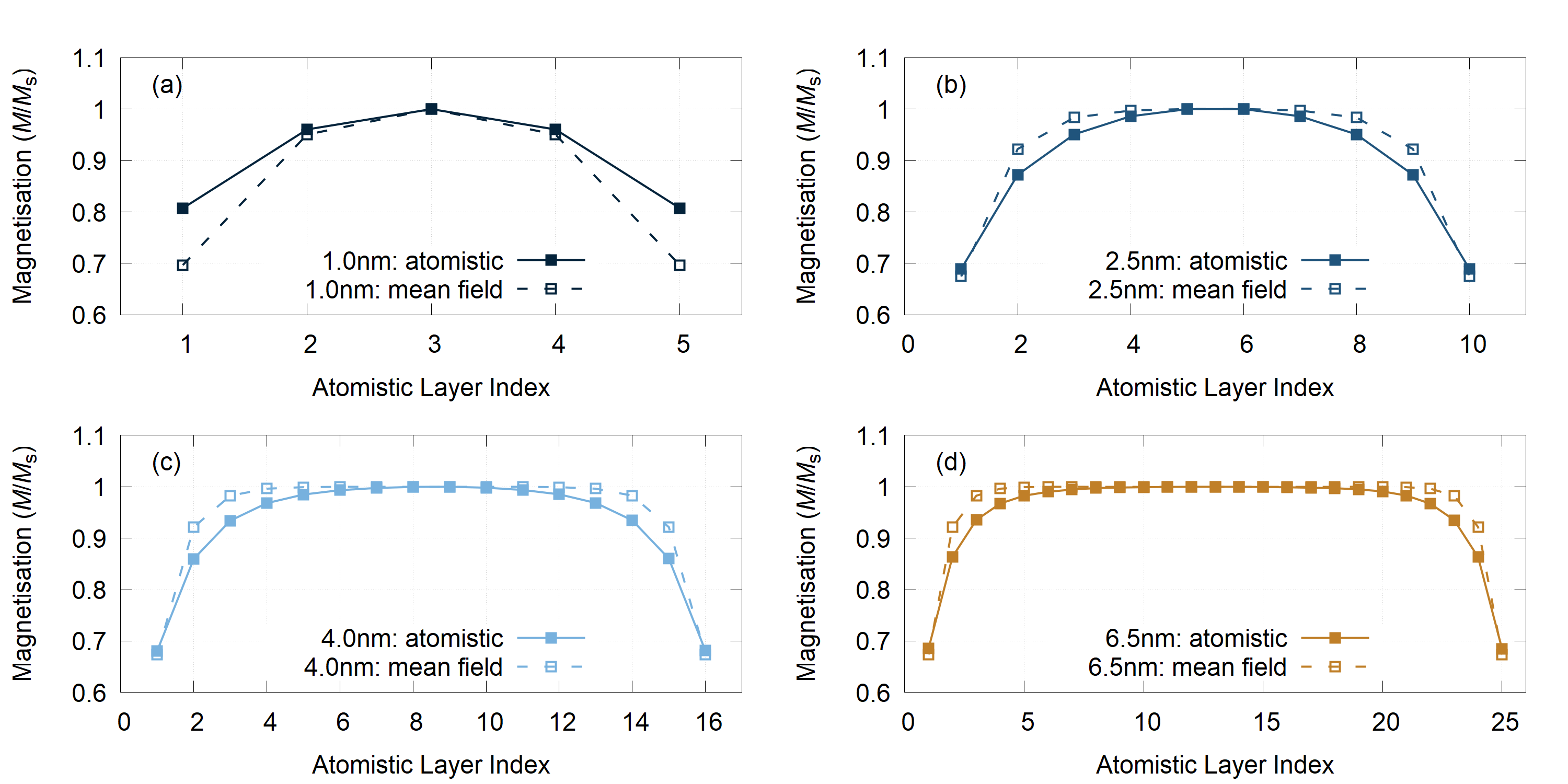}
   \caption[]{Comparison between atomistic and semi-analytic mean field calculation of layer-resolved magnetization at 550K for grains of size (a) 1.0nm, (b) 2.5nm, (c) 4.0nm, and (d) 6.5nm. A good agreement is shown between 2 models, with a slight disparity occurring at the outermost surface layers of each grain.}
\label{fig_Meanfield}
\end{figure*}

The results obtained from atomistic simulations are next compared with calculations from the Mean-field model described in the theory. Two regimes of behavior for small and large grain size can be seen in Fig.~\ref{fig_Meanfield}. Particularly, the larger grains retain order in the central region, with increasing loss of order close to the surface, whereas for the smaller grains the disorder essentially penetrates the whole grain. The mean-field model gives good qualitative agreement with the atomistic model calculations, supporting the localization of the disorder close to the surface of the grain. In the following we present a simple analysis designed to characterize  the penetration depth of the disordered region. 

\begin{figure*}[!tb]
\centering
 \includegraphics[angle = -0,width = 1.0 \linewidth]{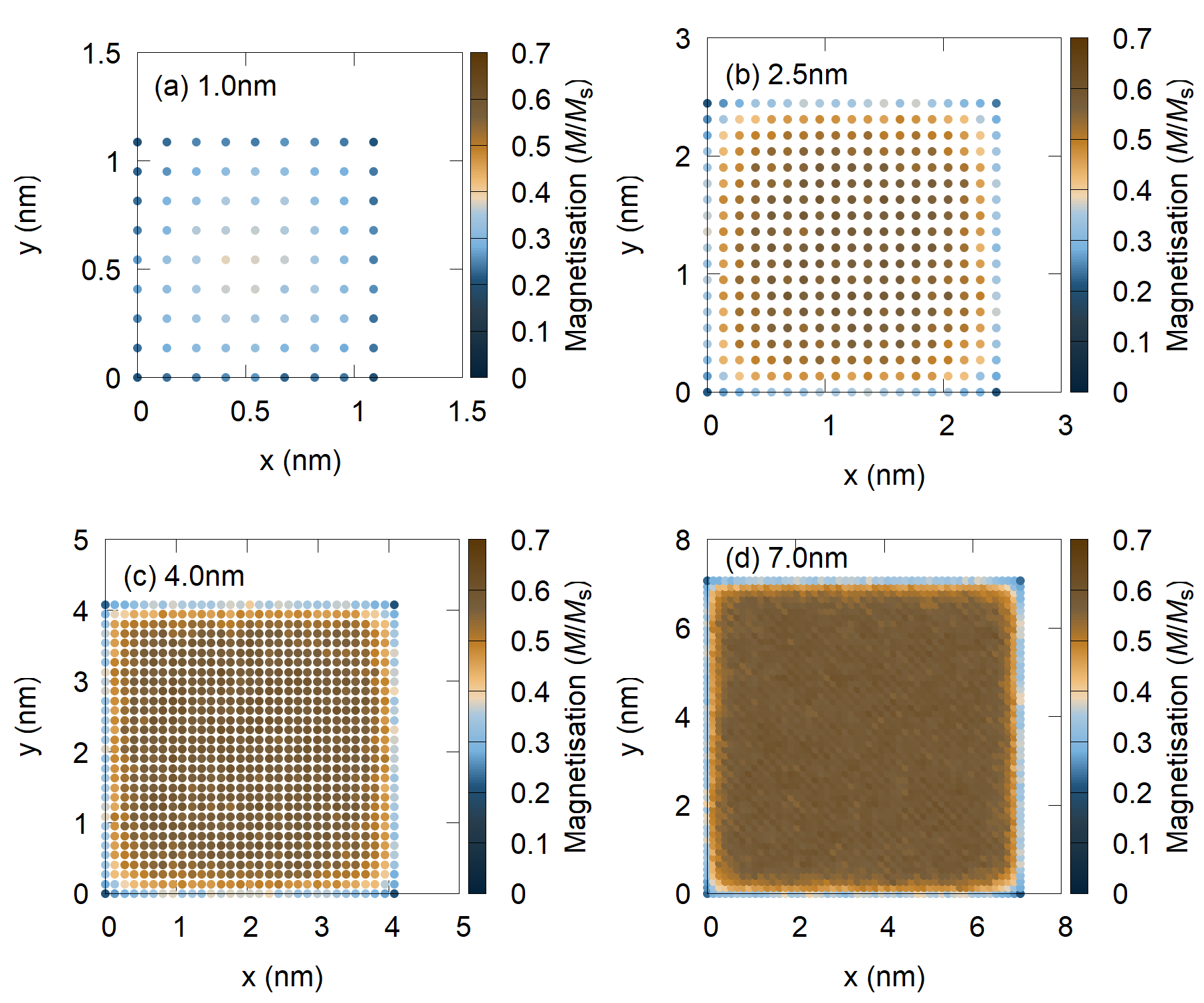}
\caption{Evolution of the cross-sectional magnetization profile for fcc lattice grains at 550K of (a) 1.0nm, (b) 2.5nm, (c) 4.0nm, and (d) 7.0nm in size. The grain magnetization can be seen to be decreasing across the grain surface.}
\label{fig_Heatmap-fcc}
\end{figure*}

The evolution of the cross-sectional magnetization profile for fcc lattice grains at 550K in Fig.~\ref{fig_Heatmap-fcc} shows a decrease of magnetization across the grain surface which appears to be more pronounced in smaller grains, consistent with the data shown in Fig.~\ref{fig_MagnetisationProfile-fcc}. Results for modified-sc and bcc lattices are given in the Supplemental section S3. These patterns suggest that surface disorder might be an important contribution to the rapid drop in $T_C$ at smaller sizes as captured before by the FSSL: the hypothesis is that the propagation of the surface disorder into the grain has an effect on the correlation length such that $\xi=\xi(D)$. This hypothesis is consistent with the effects of a term  correspond to renormalization of state energies resulting from fluctuations~\cite{goldenfeld2018lectures} thereby modifying the correlation length. However, in Fig.~\ref{fig_MagnetisationProfile-fcc} it is clear  that the renormalization of the state energies decreases with distance away from the surface.  

\begin{figure}[ht!]
\centering
 \includegraphics[angle = -0,width = 1.0 
 \linewidth]{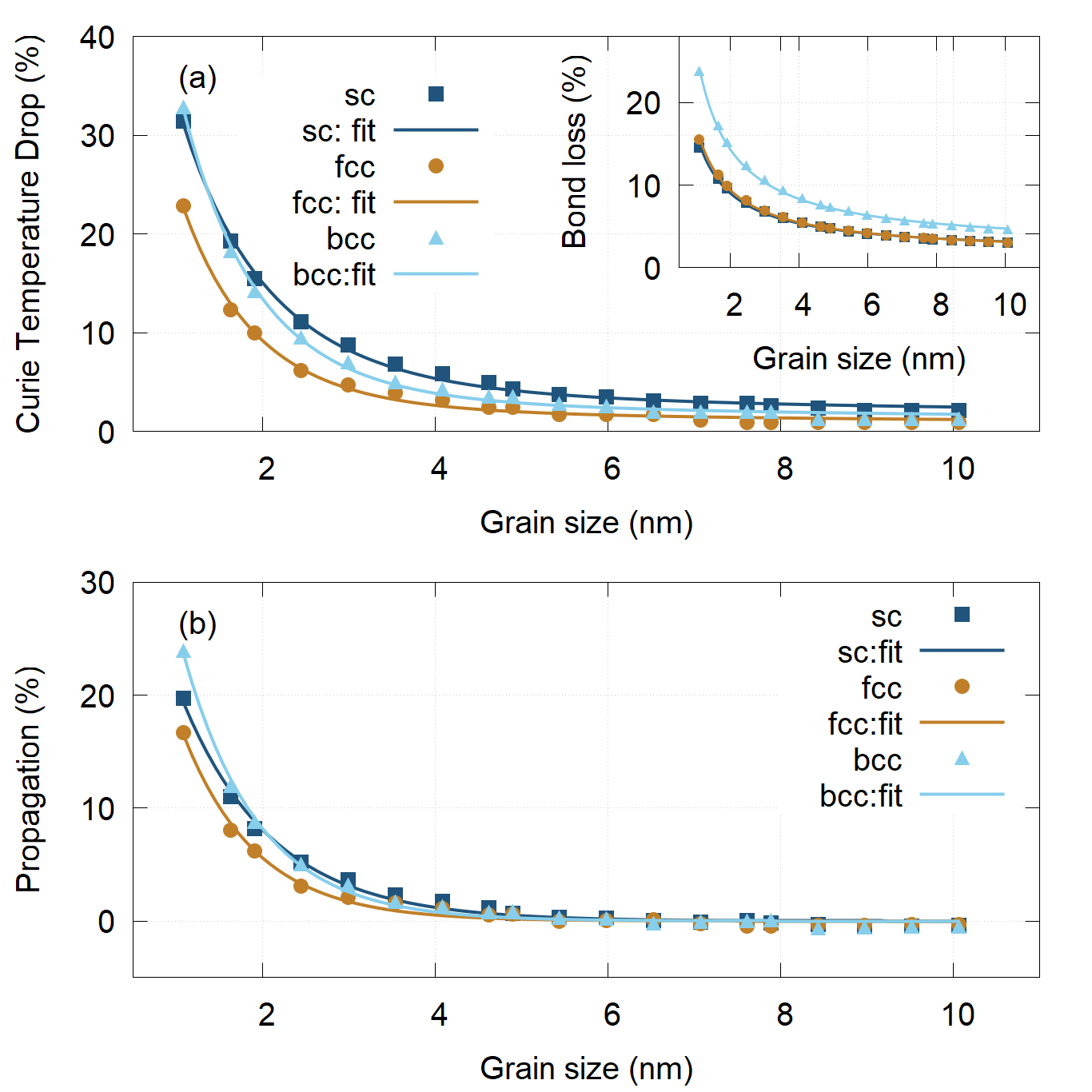}
   \caption[]{Correlation between the Curie temperature drop and atomistic bond loss for each lattice structure: (a) ${\Delta}T_{C}$(D) following from the FSSL fit and the inset showing the percentage bond loss; (b) The surface-effects propagation term ${\Delta}T_{cp}$ showing a cut-off value at $D\approx$3.5nm for all 3 studied lattice structures.}
\label{fig_Correlation}
\end{figure}

We now develop a simple analysis reflecting both contributions. The starting point is to assume that the Curie temperature reduction is entirely due to the loss of coordination at the surface. Thus $\Delta T_c$ is assumed proportional to the number of surface bonds broken. As a first approximation we assume that the number of broken bonds $n_{bb}$ is proportional to the surface area of the grains. Then it is straightforward to show that the fractional increase in broken bonds $\Delta n_{bb}$ (relative to the number of bonds in the bulk) as a function of diameter is given by:
\be
   \Delta n_{bb}=B(h^{-1} +2D^{-1}),
   \label{Eq:bondslost}
\ee
where $h$ is the height of the grain and $B$ is a constant depending on the crystal structure and is determined by fitting to the numerical calculation. Eq.~\eqref{Eq:bondslost} gives a good fit to the numerical results for all lattice structures studied, as shown in the inset of Fig.~\ref{fig_Correlation}(a). In terms of fitting to the values of $\Delta T_c$, we find that the assumption of $\Delta T_c \propto \Delta n_{bb}$ with $\Delta n_{bb}$ following the expression in Eq.~\eqref{Eq:bondslost} is valid only for large diameters, indicating that in this regime the decrease in $T_C$ is essentially a surface effect. As we show later, for small diameter the surface disorder  propagates into the center of the grain leading to a more rapid decrease of $T_C$, albeit one which is captured by the finite size scaling law. We quantify this by fitting to a modified function:
\be
    \Delta T_c = \alpha \exp (-D/D_p)+\beta \Delta n_{bb},
    \label{equ:propagation}
\ee
where $\alpha$ and $\beta$ are fitting constants and $D_p$ is a characteristic distance associated with the propagation of the disorder into the center of the grain. As shown in Fig.~\ref{fig_Correlation}(a), Eq.~\eqref{equ:propagation} fits well to the numerical calculation for all lattice structures. Values of the fitting constants are given in Table~\ref{tab_Correlation}.

We define a propagation term, representing the propagation of surface disorder into the grain as follows:
\be
    \Delta T_{cp} = \Delta T_c -\beta \Delta n_{bb}.
    \label{equ:propagation2}
\ee
Equation Eq.~\eqref{equ:propagation2}, along with the numerical calculation is shown in Fig.~\ref{fig_Correlation}(b). It can be seen that, for diameter $D\lessapprox 3.5$nm there is an exponential increase in $\Delta T_c $ as the diameter decreases. Clearly the surface bonds lost drive the loss of magnetic order and the reduction in $T_C$. However, it is important to note the role played by the renormalization of the state energies arising from the fluctuations originating at the surface. This gives rise to the progressive decrease of the loss of magnetization when moving toward the center of the grain. For small grain sizes $D\le D_p$ with $D_p$ the penetration depth from Eq.~\eqref{equ:propagation}, the decrease in magnetic order due to state energy renormalization cannot be stabilized by a fully-ordered central core. In this regime the state energy renormalization becomes the dominant factor leading to a rapid collapse of the magnetization and $T_C$. Hovorka et al.~\cite{Hovorka} have given an expression relating the dispersion of $T_C$ directly to the dispersion of diameter. This suggests that for decreasing grain sizes, such as expected for the evolution of HAMR, any grain size variation would give an increasingly large distribution of $T_C$ which could become a limiting factor for the technology. On the other hand, numerous designs for HAMR media involve coupling layers with high anisotropy and low $T_C$ with layers of lower anisotropy and higher $T_C$. It is likely, from the analysis presented here, that for strongly exchange coupled layers the surface disorder and hence reduction of the $T_C$ of the high anisotropy could be somewhat mediated by the proximity effect of a higher $T_C$ layer. 

\begin{center}
\begin{table}[ht!]
\centering
\fontsize{8}{6}
\begin{tabular}{c|llll}
Configuration & $\alpha$ & $D_{p}$ & $\beta$
\\ \hline \hline 
modified-sc & $54.4 \pm 2.7$  &  $1.049 \pm 0.046$ & $0.779 \pm 0.034$  \\
fcc & $59.4 \pm 4.3$ & $0.846 \pm 0.046$ & $0.386 \pm 0.031$  \\
bcc & $83.1 \pm 4.4$ & $0.863 \pm 0.035$ & $0.368 \pm 0.022$ \\ 
\hline
\end{tabular}
\caption[]{Correlation fitting parameters for each simulated lattice configuration of FePt.}
\label{tab_Correlation}
\end{table}
\end{center}
%

\section{CONCLUSION}
\label{S:4}
We have investigated finite-size effects in small grains. Simulation data fit to the classic Finite-size Scaling Law and show a rapid decrease of $T_C$ at small sizes. We show that this is due to the propagation of surface disorder resulting from the loss of exchange coordination at the surface into the grain. This effect becomes important at grain sizes smaller than 4nm and is supported by semi-analytic Mean-field calculation. Our findings overall are consistent with the Mean-field calculation and have been extended to incorporate different crystal structures, which strongly suggests that if using a suitable correlation factor the $T_C$ distribution of a generic material can be correlated to the percentage of atomistic bond loss on the surface as a universal parameter. The reduction of $T_C$ is driven by surface magnetic disorder resulting from the loss of exchange coordination at the surface. These fluctuations cause a renormalization of state energies through which the magnetic disorder propagates into the grain. A physically reasonable expression is proposed which separates the two processes, and defines a penetration depth $D_p$ for the propagation of disorder into the grain. For small grain sizes less than around 3-4nm, the decrease in magnetic order due to state energy renormalization cannot be stabilised by a fully-ordered central core. In this regime the state energy renormalization becomes the dominant factor leading to a rapid collapse of the magnetization and $T_C$ and a consequent increase of the dispersion of $T_C$ for small diameter. From the viewpoint of materials design for nanoscale applications such as spintronics and particularly Heat-assisted Magnetic Recording, finite-size effects will become an increasingly important consideration with decreasing device size. Because of the strong surface effects on the decrease of $T_C$, the increased  $T_C$ dispersion for small grains could be somewhat mediated in designs coupling low $T_C$ hard materials such as FePt with high $T_C$ materials which would reduce the loss of magnetic order through the proximity effect.

\bibliography{reference}
\end{document}


\title{The influence of finite size effects on the Curie temperature of L1\textsubscript{0}-FePt: Supplemental information}
\author{Nguyen Thanh Binh}
\affiliation{%
Department of Physics, the University of York, Heslington, York YO10 5DD, United Kingdom}%
\author{Sergiu Ruta}
\affiliation{%
Department of Physics, the University of York, Heslington, York YO10 5DD, United Kingdom}%
\author{Ondrej Hovorka}
\affiliation{%
Faculty of Engineering and Physical Sciences, the University of Southampton, Southampton
SO17 1BJ, United Kingdom}%
\author{Richard. F. L. Evans}
\affiliation{%
Department of Physics, the University of York, Heslington, York YO10 5DD, United Kingdom}%
\author{Roy. W. Chantrell}
\affiliation{%
Department of Physics, the University of York, Heslington, York YO10 5DD, United Kingdom}%
\maketitle
\section{S1: Mean-field Theory Calculation of the Curie Temperature}
Here we show that Eq.~(7) (main text) can be derived from Eq.~(6) (main text). First write Eq.~(6) in the scalar form as:
\be
    m_i = \mathcal{L}(\beta\mu B^e_i),     
\ee 
where:
\be
\label{beff}
\begin{split}
    \mu B^e_i = J\sum_{j \in i} m_j + \mu B.
\end{split}
\ee
Assume zero applied field $B=0$ and express $m_i = \mathcal{L}(\beta J\sum_j m_j)$. Imagine the system is in a paramagnetic state, then $m_i=0$ for all $i$ in the Mean-field model. When the freezing occurs at $T_c$ then all spins will freeze co-linearly (because there is no DMI which would prefer non-collinear spin alignment, for example), and get some finite moment $m_i=m$. Rewrite the equation as $m = \mathcal{L}(\beta Jz m)$, where $z$ is the spin-coordination number, i.e. $\sum_j m_j = zm$. Differentiate with respect to the field (imagine an increment to small non-zero field), which gives:
\be
    \frac{dm}{dB} = \mathcal{L}'.\left(\beta Jz\right).\frac{dm}{dB}, 
\ee
where $\mathcal{L}'$ is the derivative of the Langevin function $\mathcal{L}$, which expanded to the first order gives us the factor $1/3$. Put this in and express:
\be
    \frac{dm}{dB}\left(1-\frac{1}{3}\beta Jz\right) = 0.
\ee
This has a non-trivial solution if $3^{-1}\beta Jz=1$, which upon arranging gives:
\be
    T_c=\frac{zJ}{3k_B}.
\ee
This is the same as  Eq.(7) (main text), apart from the constant $\epsilon$ which in the mean-field model $\epsilon=1$. Thus we are essentially using a mean-field formula with a correction factor $\epsilon$, which is needed in order to `fit' the mean-field solution ansatz to non-Mean-field models. 
\newpage
%
\section{S2: Layer-resolved Magnetization Profiles for sc and bcc Lattices}
\begin{figure}[bth!]
\centering
\begin{subfigure}[a]{0.45\textwidth}
\includegraphics[width=\linewidth]{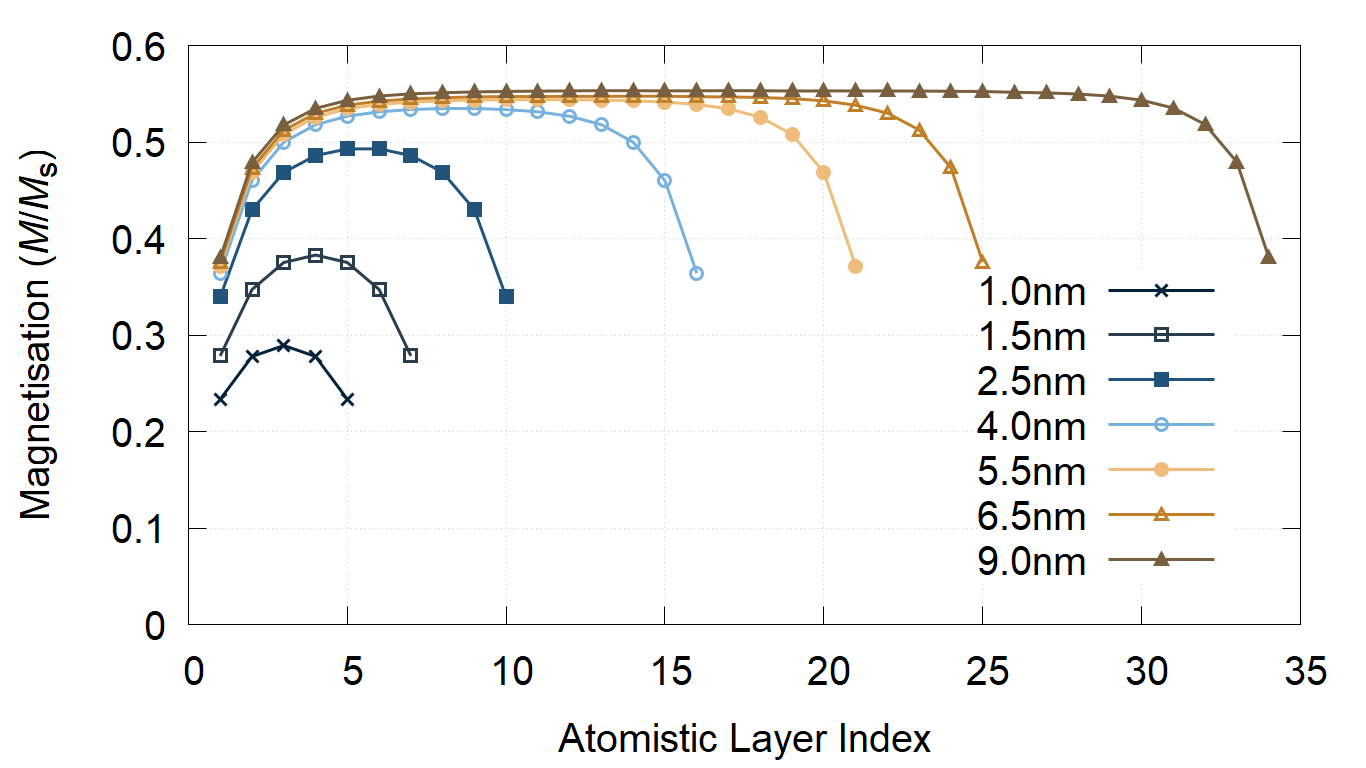}
    \caption{ }
    \end{subfigure}\hfill
    \begin{subfigure}[b]{0.45\textwidth}
\includegraphics[width=\linewidth]{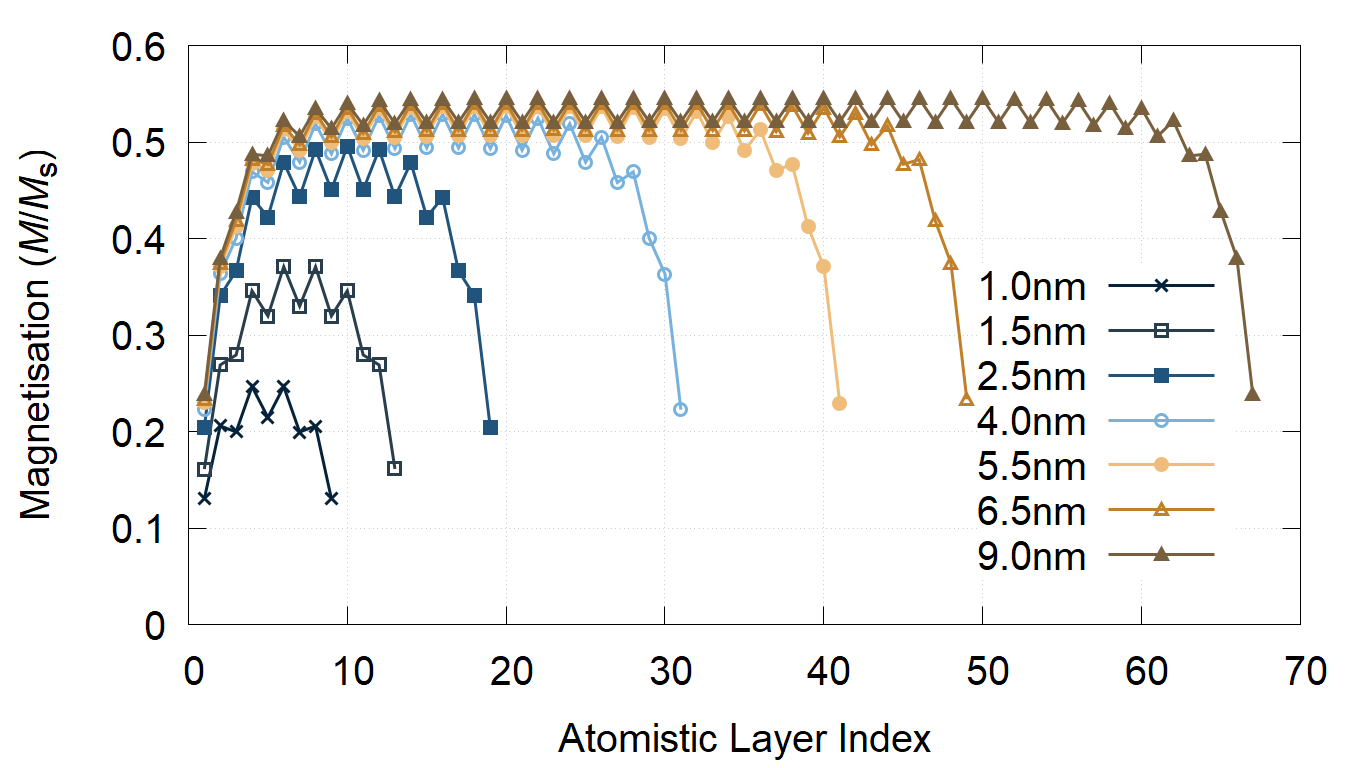}
    \caption{}
    \end{subfigure}
\caption{The layer-resolved magnetization profile at 550K for (a) modified-sc and (b) bcc lattice grains. The bcc lattice structure exhibits a periodic behavior which is a physical effect due to atoms having different numbers of nearest neighbours.}
\label{fig_MagnetisationProfiles}
\end{figure} 

Here we show the magnetization profiles for modified-sc - Fig.~\ref{fig_MagnetisationProfiles}(a) - and bcc - Fig.~\ref{fig_MagnetisationProfiles}(b) - lattices. As expected the behavior is similar to the fcc results given in the main text, although the penetration depth of clearly smaller for the modified-sc lattice. It can be seen that there is a periodic behavior for the bcc lattice, which is a physical effect arising from atoms having different numbers of nearest neighbors. Interestingly this persists to elevated temperatures in the bulk of the grain albeit somewhat reduced at the grain boundaries, suggesting that the disorder propagates inward from the surfaces.
\newpage
%

\section{S3: Cross-sectional Magnetization Profiles for sc and bcc Lattices}
\begin{figure}[bth!]
\centering
\begin{subfigure}[a]{0.45\textwidth}
\includegraphics[width=\linewidth]{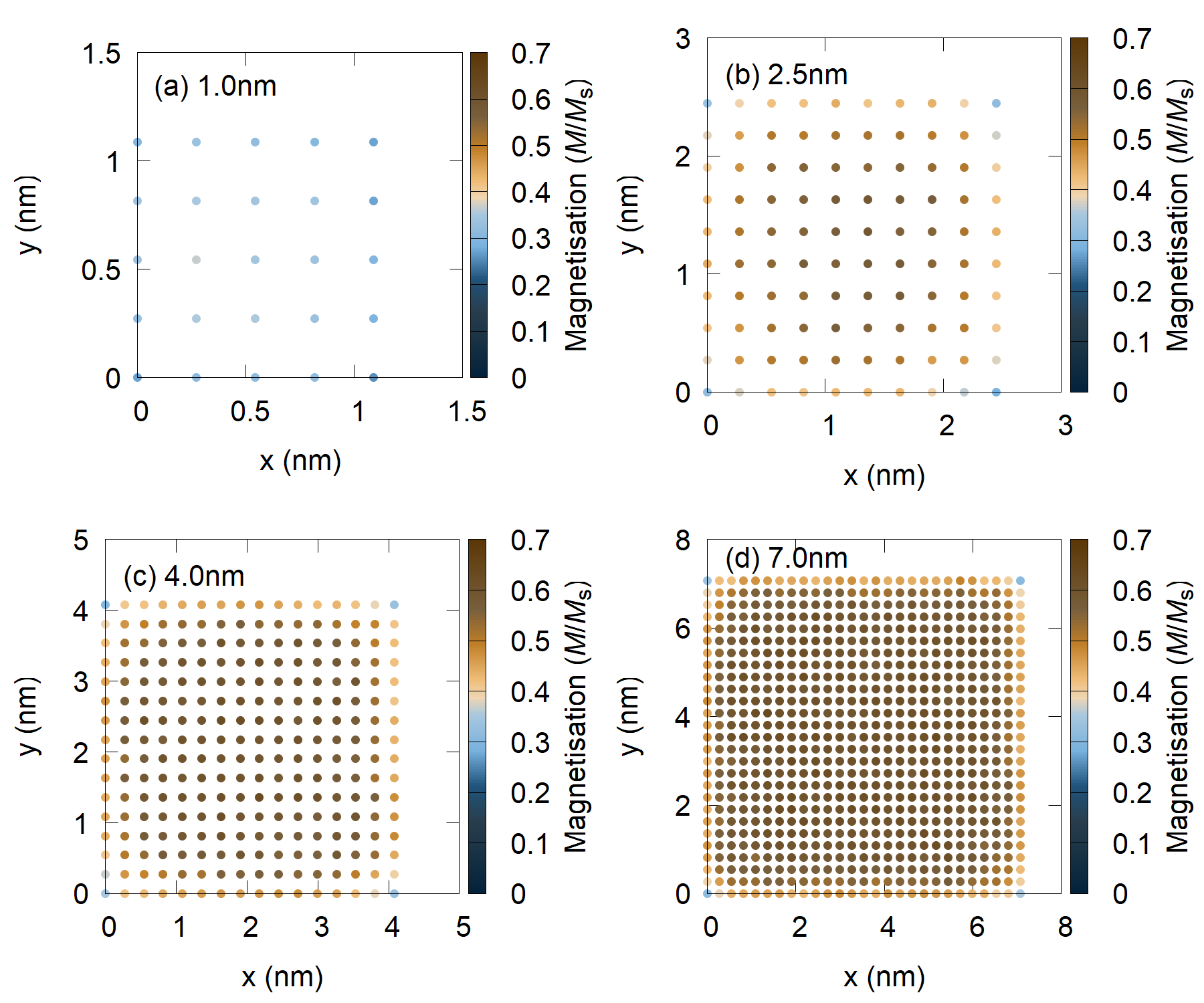}
    \caption{ }
    \end{subfigure}\hfill
    \begin{subfigure}[b]{0.45\textwidth}
\includegraphics[width=\linewidth]{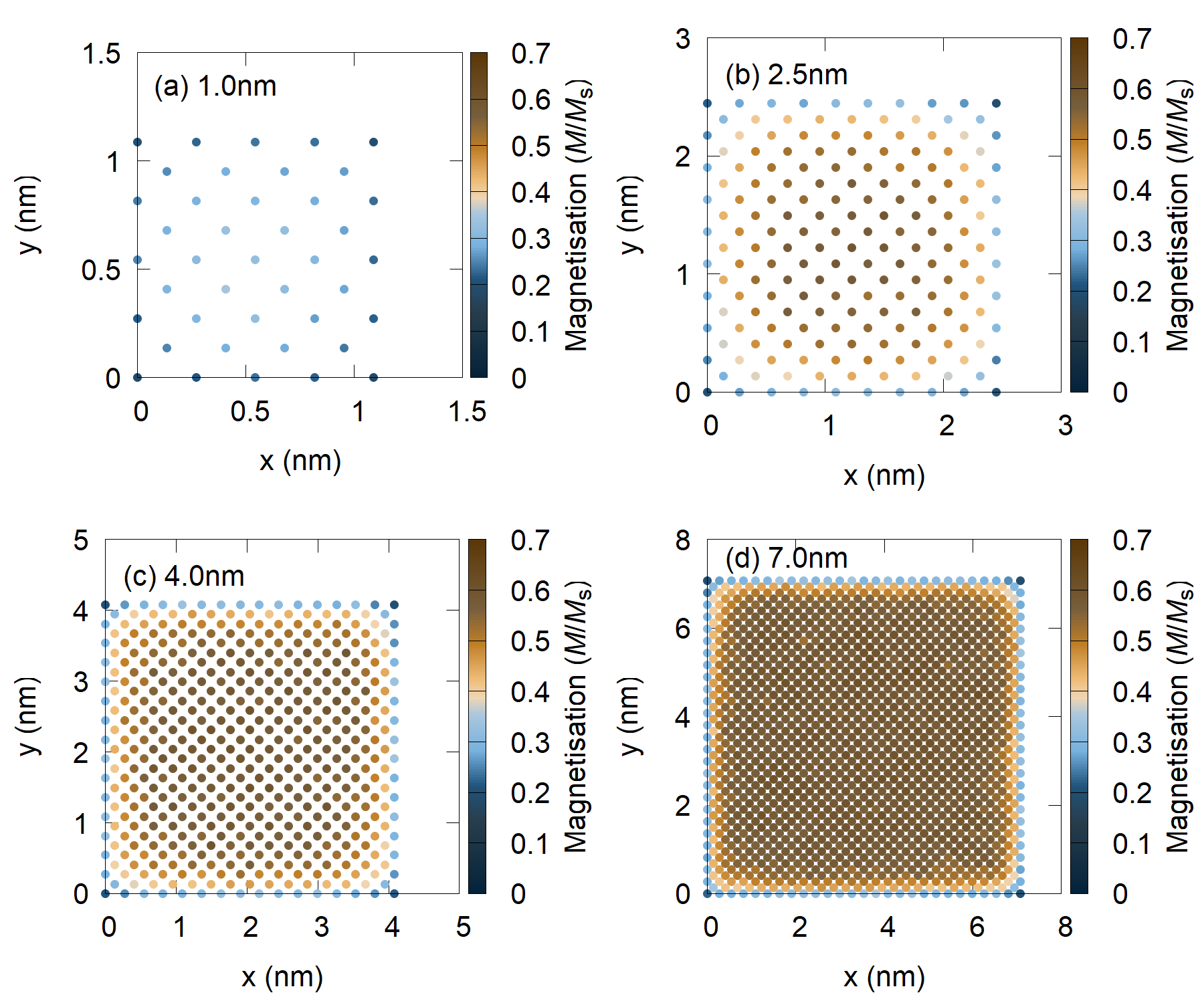}
    \caption{}
    \end{subfigure}
\caption{Evolution of the cross-sectional magnetization profile at 550K for (a) modified-sc and (b) bcc grains of 1.0nm, 2.5nm, 4.0nm, and 7.0nm in size. The same behaviours are observed as for fcc lattice grain with the magnetization decreasing across the grain surface.}
\label{fig_MagnetisationHeatmaps}
\end{figure}